\documentclass[fleqn,usenatbib]{mnras}

\usepackage{newtxtext,newtxmath}
\usepackage[T1]{fontenc}
\usepackage{ae,aecompl}

\usepackage{graphicx}	% Including figure files
\usepackage{amsmath}	% Advanced maths commands
\defcitealias{mds19}{Paper I}
\defcitealias{santos+20}{Paper II}
\defcitealias{dias+21}{Paper III}
\defcitealias{dias+22}{Paper IV}

\title[Rejuvenating three faint SMC clusters]{The VISCACHA survey -- V. Rejuvenating three faint SMC clusters}

\author[E. Bica et al.]
{E. Bica$^{1}$\thanks{E-mail: bica@if.ufrgs.br},
F. F. S. Maia$^{2}$,
R. A. P. Oliveira$^{3}$,
B. Dias$^{4}$,
J. F. C. Santos Jr.$^{5}$,
J. P. Rocha$^{6}$,
\newauthor
L. Kerber$^{6}$,
J. F. Gardin$^{5}$,
T. Armond$^{7}$, 
M. C. Parisi$^{8,9}$,
S. O. Souza$^{10,3}$,
B. Barbuy$^{3}$
\\
$^{1}$Departamento de Astronomia, IF - UFRGS, Av. Bento Gon\c calves 9500, 91501-970, Brazil \\ 
$^{2}$Instituto de F\'isica - Universidade Federal do Rio de Janeiro, Av. Athos da Silveira Ramos, 149, Rio de Janeiro, 21941-909, Brazil \\
$^{3}$Universidade de S\~ao Paulo, IAG, Rua do Mat\~ao 1226, Cidade Universit\'aria, S\~ao Paulo 05508-900, Brazil\\
$^{4}$Instituto de Alta Investigaci\'on, Sede Esmeralda, Universidad de Tarapac\'a, Av. Luis Emilio Recabarren 2477, Iquique, Chile\\
$^{5}$Departamento de F\'isica, ICEx - UFMG, Av. Ant\^onio Carlos 6627, Belo Horizonte 31270-901, Brazil \\ 
$^{6}$Departamento de Ci\^encias Exatas e Tecnol\'ogicas, UESC, Rodovia Jorge Amado km 16, 45662-900, Brazil\\
$^{7}$Universidade Federal de São João del-Rei, Departamento de Estatística, Física e Matemática, Campus Alto Paraopeba, Rod.: MG 443, Km 7, \\~~~Ouro Branco - MG, 36420-000, Brazil\\
$^{8}$Observatorio Astron\'omico, Universidad Nacional de C\'ordoba, Laprida 854, X5000BGR, C\'ordoba, Argentina\\
$^{9}$Instituto de Astronom{\'\i}a Te\'orica y Experimental (CONICET-UNC), Laprida 854, X5000BGR, C\'ordoba, Argentina\\
$^{10}$Leibniz-Institut für Astrophysik Potsdam (AIP), An der Sternwarte 16, Potsdam, 14482, Germany}

\date{Accepted XXX. Received YYY; in original form ZZZ}

\pubyear{2022}

\begin{document}
\label{firstpage}
\pagerange{\pageref{firstpage}--\pageref{lastpage}}
\maketitle

\begin{abstract}
  We present the analysis of three faint clusters of the Small Magellanic Cloud RZ\,82, HW\,42 and RZ\,158. We employed the SOAR telescope instrument SAM with adaptive optics, allowing us to reach to  $V\sim 23-24$\,mag, unprecedentedly, a depth sufficient to measure ages of up to about 10-12 Gyr. All three clusters are resolved to their centres, and the resulting colour-magnitude diagrams (CMDs) allow us to derive ages of 3.9, 2.6, and 4.8\,Gyr respectively. These results are significantly younger than previous determinations (7.1, 5.0, and 8.3\,Gyr, respectively), based on integrated photometry or shallower CMDs. We rule out older ages for these clusters based on deep photometry and statistical isochrone fitting. We also estimate metallicities for the three clusters of $\rm{[Fe/H]}=-0.68$, $-0.57$ and $-0.90$, respectively. These updated ages and metallicities are in good agreement with the age-metallicity relation for the bulk of SMC clusters. Total cluster masses ranging from $\sim 7-11\cdot10^3\,M_\odot$ were estimated from integrated flux, consistent with masses estimated for other SMC clusters of similar ages. These results reduce the number of SMC clusters known to be older than about 5 Gyr and highlight the need of deep and spatially resolved photometry to determine accurate ages for older, low-luminosity SMC star clusters.
\end{abstract}


%%%%%%%%%%%%%%%%%%%%%%%%%%%%%%%%%%%%%%%%%%%%%%%%%%

%%%%%%%%%%%%%%%%% BODY OF PAPER %%%%%%%%%%%%%%%%%%

\section{Introduction}
\label{sec:intro}

The Small Magellanic Cloud (SMC) cluster population is a tracer of the star and cluster formation history and chemical enrichment in the SMC \citep[e.g.][]{ppv17}. The SMC cluster population is also important to help understanding the interactions with the Large Magellanic Cloud (LMC) that seem to have caused some bursts of cluster formation \citep[e.g.][]{piatti+11}. The information on the early SMC evolution since the SMC-LMC interactions started to happen is not yet well constrained because there is a lack of SMC clusters older than $\sim 8$\,Gyr \citep{p11c, parisi+14}. Therefore, it is of prime importance to derive accurate ages for candidates to be the oldest star clusters in the SMC.

\citet{csp01} showed with \textit{HST} data that the populous old clusters in the SMC span ages from 1\,Gyr to the NGC\,121 age \citep[$10.5-11.8$\,Gyr,][]{glatt+08}, which is comparable to the ages of Galactic globular clusters. NGC\,121 is the only known massive cluster older than 8\,Gyr, but the SMC contains a much larger population of intermediate to low mass clusters to be explored \citep[e.g.][]{gatto+21}. Attempts to derive ages from ground-based data \citep[e.g.][]{ggk10} were hampered by the photometric limit hardly attaining the main sequence turnoff (MSTO).

The VISCACHA \citep[VIsible Soar photometry of star Clusters in tApii and Coxi HuguA;][hereafter \citetalias{mds19}]{mds19} survey is an ongoing project that employs the 4.1-m Southern Astrophysical Research (SOAR) telescope aided by adaptive optics that provides photometry deeper than the MSTO of old Magellanic Clouds clusters and resolved stars in the cluster cores. The VISCACHA data are designed to derive accurate ages and masses of the oldest intermediate to low mass clusters of the SMC.

\begin{table*}
	\centering
	\caption{Log of observations for the three analysed clusters.}
	\label{tab:obslog}
	\begin{tabular}{lccccccccc} 
		\hline
Cluster & RA (J2000) & Dec. (J2000) & Date & Instrument & Filters & Exp. time & Seeing & FWHM & Airmass \\
 & (h:m:s) & ($^\circ$:$^\prime$:$^{\prime\prime}$) &  &  &  & (s) & (arcsec) & (arcsec) & \\
		\hline
RZ\,82   & 00:53:09.6 & -71:59:43 & 2018/12/12 & SAMI & V, I & 3x400, 3x600 & 0.80 & 0.81, 0.69 & 1.38 \\
HW\,42   & 01:01:06.3 & -74:04:32 & 2018/10/06 & SAMI & V, I & 3x400, 3x600 & 0.85 & 0.90, 0.63 & 1.53 \\
RZ\,158  & 01:06:45.0 & -74:49:58 & 2021/11/08 & SAMI & V, I & 3x400, 3x600  & 0.60 & 0.72, 0.63 & 1.45 \\
\hline
	\end{tabular}
\end{table*}

\begin{figure*}
  \includegraphics[width=0.32\textwidth]{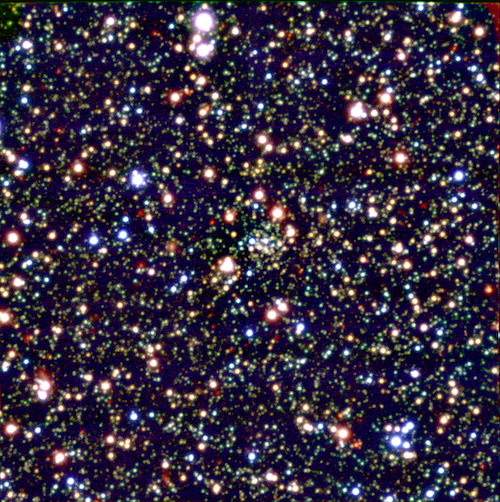}
  \includegraphics[width=0.32\textwidth]{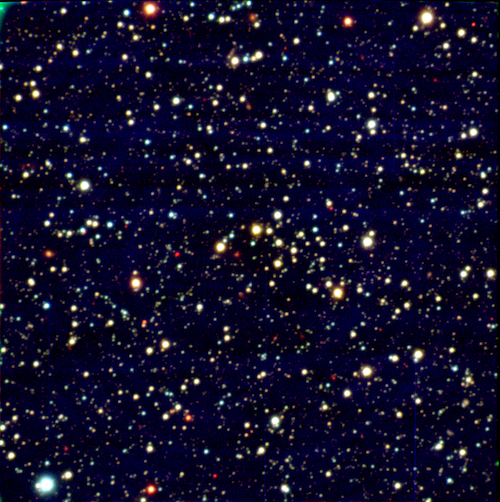}
  \includegraphics[width=0.32\textwidth]{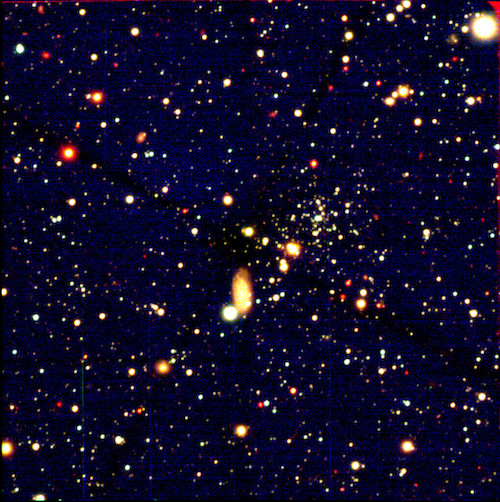}

  \caption{Colour composite images of RZ\,82 (left), HW\,42 (middle) and RZ\,158 (right) made from SAMI exposures. The images have $3^\prime\times 3^\prime$, corresponding to the SAMI FoV. North is up and East to the left.}
    \label{fig:ima}
\end{figure*}

In this work we analyse three old cluster candidates from \citet{bica+20} catalogue. Clusters RZ\,82 and RZ\,158 \citep{rz05} had integrated colours suggesting an intermediate or old age. \citet{p11c} estimated an age of 9.3\,Gyr for HW\,42 with Washington photometry at Blanco telescope. \citet{ppv17} found that HW\,42 might be as old as 5\,Gyr with a total photometric mass of $200\,M_\odot$ with the same data.
The VISCACHA data for these clusters were obtained to improve on these age, metallicity and cluster mass estimates.

In Sect.~\ref{sec:obs} we present the observations and reductions, with the obtained $BVI$ images. In Sect.~\ref{sec:fits} we discuss the statistical isochrone fitting. The clusters' masses are presented in Sect.~\ref{sect:mass}. A discussion is given in Sect.~\ref{sec:dis} and the conclusions are drawn in Sect.~\ref{sec:conc}.

%%%%%%%%%%%%%%%%%%%%%%%%%%%%%%%%%%%%%%%%%%%%%%%%%%%%%%%
\section{Observations and reductions}
\label{sec:obs}

The observations were carried out with the 4.1-m telescope SOAR, employing the SOAR Adaptive Module \citep{tokovinin+16} with ground-layer adaptive optics and the associated imager (SAMI). We used standard reductions with bias subtraction and flat field correction. GSC2.3 \citep{llm08} was used for the astrometric calibration while standard fields from \cite{s00} were observed for photometric calibrations. More details are given in \citetalias{mds19}. Table~\ref{tab:obslog} summarises the observational information. 

Fig.~\ref{fig:ima} shows the combined cluster images according to Table~\ref{tab:obslog}. Fig. \ref{fig:cmd-all} shows the respective CMDs, from where it becomes clear that the faintest stars reach around $V\sim 23.5$\,mag for the three clusters, indicating an unprecedented photometry performance in terms of depth and spatial resolution, compared to literature data. Although faint, the clusters have relatively well-defined structures, with a core density that contrasts with the surroundings.

%%%%%%%%%%%%%%%%%%%%%%%%%%%%%%%%%%%%%%%%%%%%%%%%%%%%%%%  
\section{Statistical isochrone fitting}
\label{sec:fits}

An important first step to carry out a reliable isochrone fitting is the assignment of a membership probability for the observed stars, removing the most likely to belong to the field population. Since the proper motion data from \textit{Gaia} and other surveys are not deep enough for SMC clusters, we obtain the membership from a statistical analysis based on \citet{mcs10}, comparing the distance to the centre and local density of the stars in the CMD of the cluster sample (within the tidal radius) with a nearby field.

\begin{table}
	\centering
	\caption{Isochrone fitting results derived with the SIRIUS code. The values correspond to the median and $1\sigma$ level of the posterior distribution. Literature results are presented for comparison: \citet[][RZ05]{rz05}, 
	\citet[][P11]{p11c}, \citet[][P17]{ppv17}.}
	\label{tab:results}
	\begin{tabular}{llccccc} 
		\hline
Cluster & & Age & $\rm{[Fe/H]}$ &  $d$ & $A_{V}$ \\ 
  & & (Gyr) &     & (kpc) & (mag)\\
		\hline
RZ\,82 & (this work)   & $3.9^{+0.8}_{-0.8}$ & $-0.68^{+0.33}_{-0.33}$ & $51.1^{+4.5}_{-4.5}$ & $0.43^{+0.22}_{-0.22}$ \\ 
& (RZ05) & $7.1^{+0.7}_{-2.5}$ & $-0.58$ & -- & -- \\ 
\noalign{\smallskip}
HW\,42 & (this work)   & $2.6^{+0.3}_{-0.3}$ & $-0.57^{+0.37}_{-0.37}$ & $56.0^{+4.1}_{-4.1}$ & $0.26^{+0.26}_{-0.26}$ \\
& (P11) & $9.3^{+1.5}_{-1.5}$ & $-1.40^{+0.25}_{-0.25}$ & $60.3^{+2.8}_{-2.8}$ & $0.09^{+0.03}_{-0.03}$ \\
& (P17) & $5.0^{+5.0}_{-2.5}$ & $-0.88^{+0.65}_{-0.65}$ & $61.4^{+1.7}_{-1.7}$ & $0.05^{+0.02}_{-0.02}$ \\
\noalign{\smallskip}
RZ\,158 & (this work)  & $4.8^{+1.6}_{-1.3}$ & $-0.90^{+0.43}_{-0.39}$ & $54.7^{+3.5}_{-3.5}$ & $0.18^{+0.16}_{-0.12}$ \\
& (RZ05) & $8.3^{+1.7}_{-0.4}$ & $-0.58$ & -- & -- \\ 
\hline
	\end{tabular}
\end{table}

\begin{figure*}
	\includegraphics[trim={0 0.5cm 2.4cm 0},clip, height=8cm]{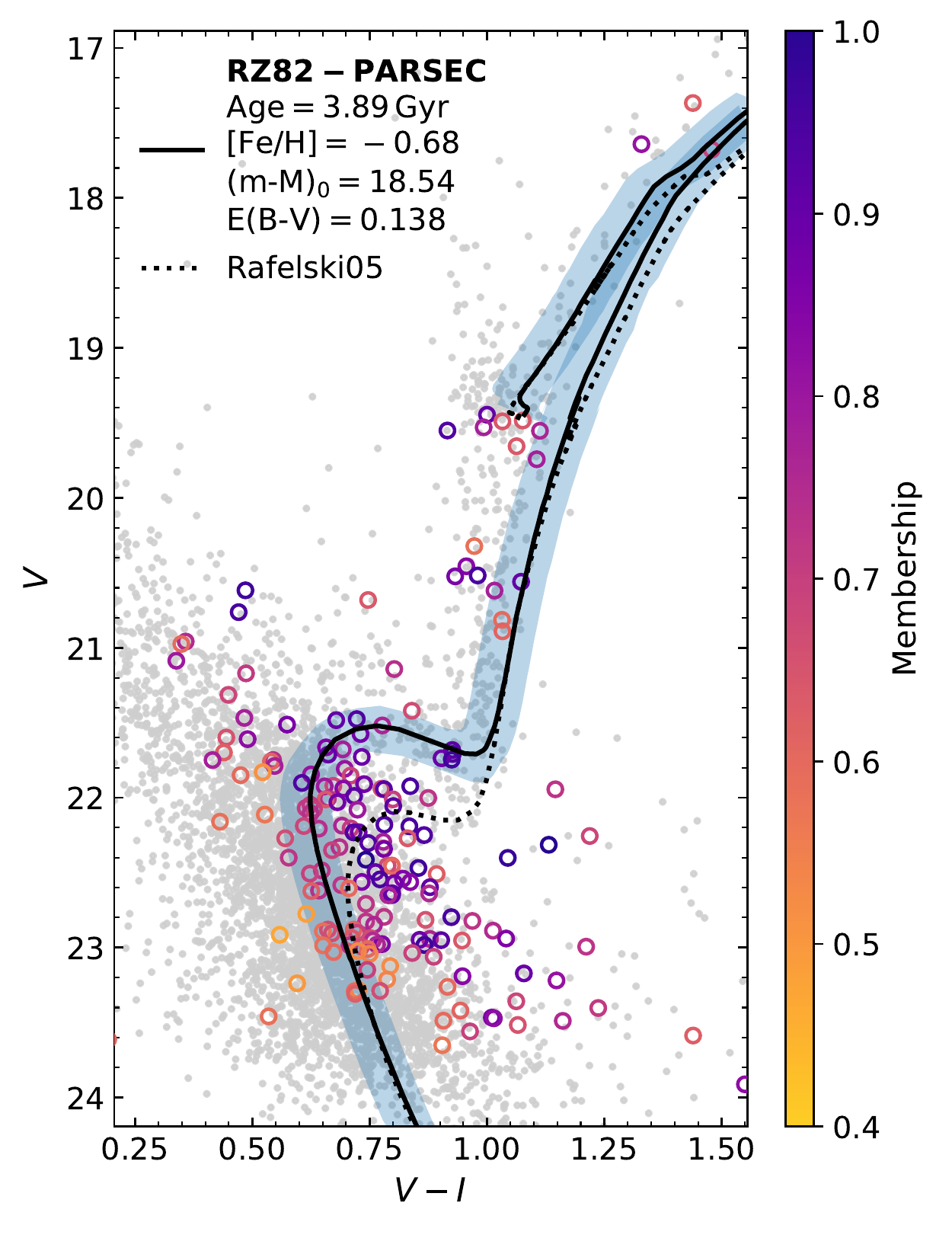}
    \includegraphics[trim={0 0.5cm 2.4cm 0},clip, height=8cm]{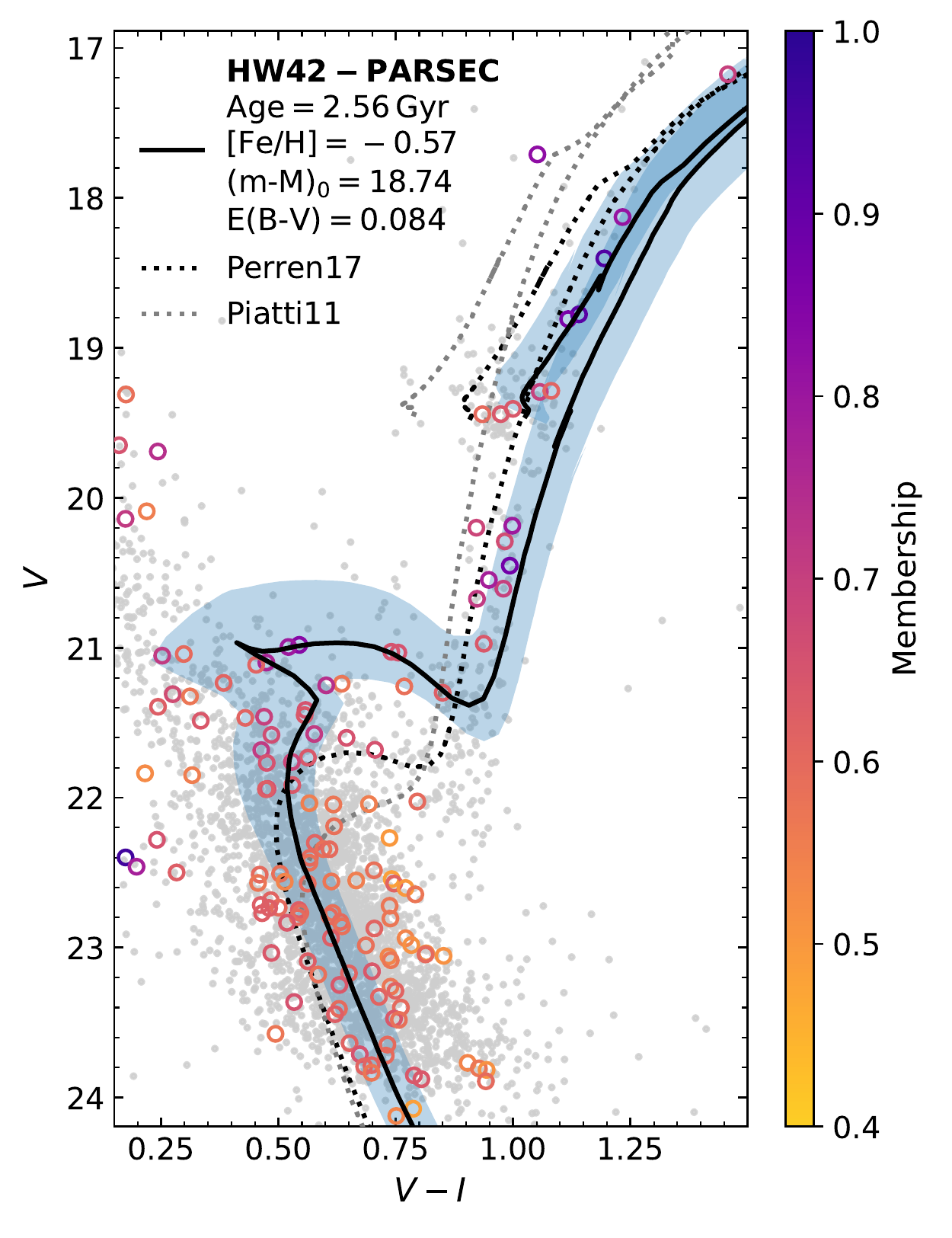}
	\includegraphics[trim={0 0.5cm 0 0},clip, height=8cm]{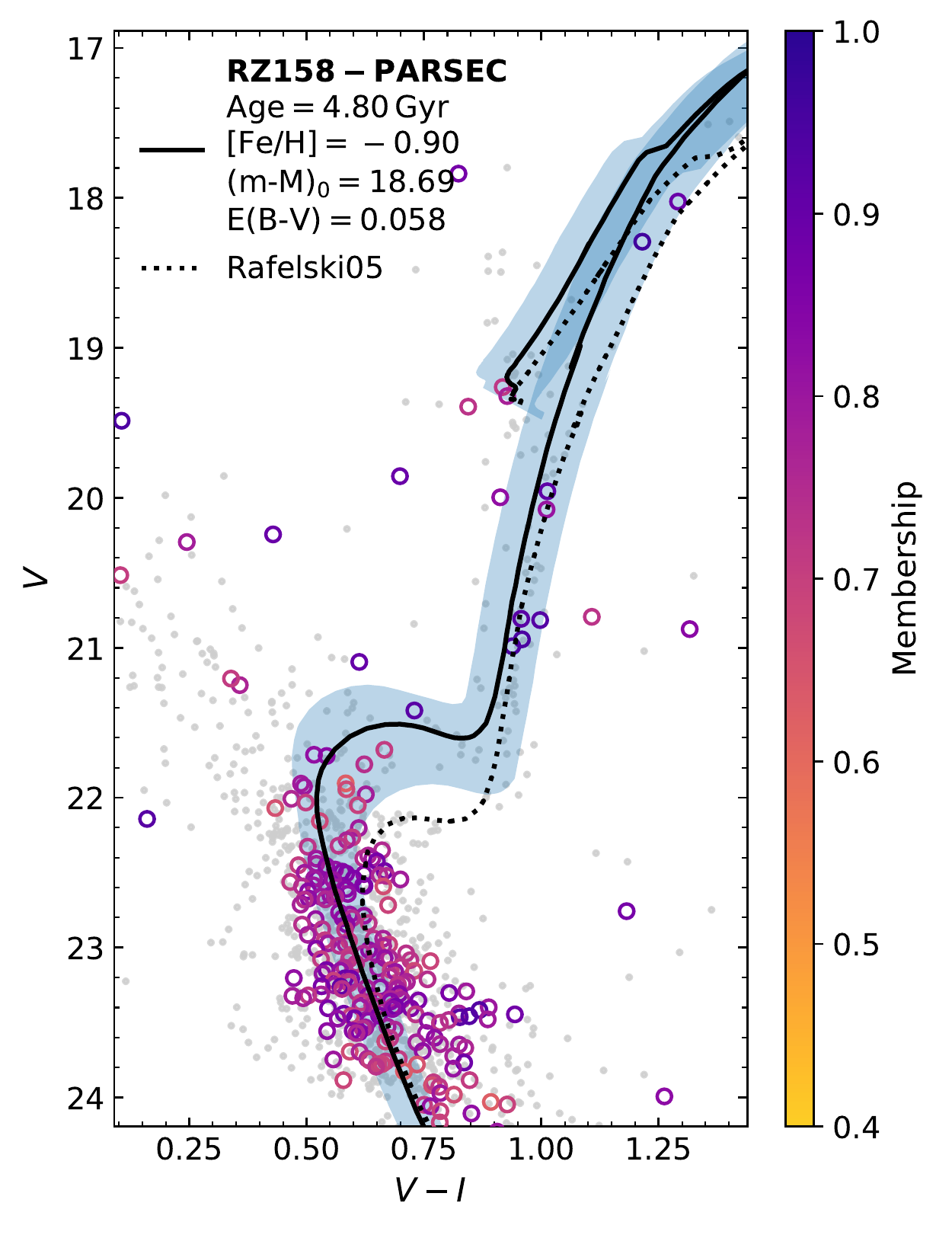}
    \caption{Decontaminated $V$ vs. $V-I$ CMDs with the isochrone fitting results for the three sample clusters. The best-fit isochrone is shown as a solid line, whereas the surrounding blue region covers all the possible solutions within $1\sigma$. The dashed isochrones give a comparison with previous literature results (see Table~\ref{tab:results}), assuming our derived distances and a suitable reddening when not available.}
    \label{fig:cmd-all}
\end{figure*}

\begin{figure*}
    \centering
    \includegraphics[width=0.33\textwidth]{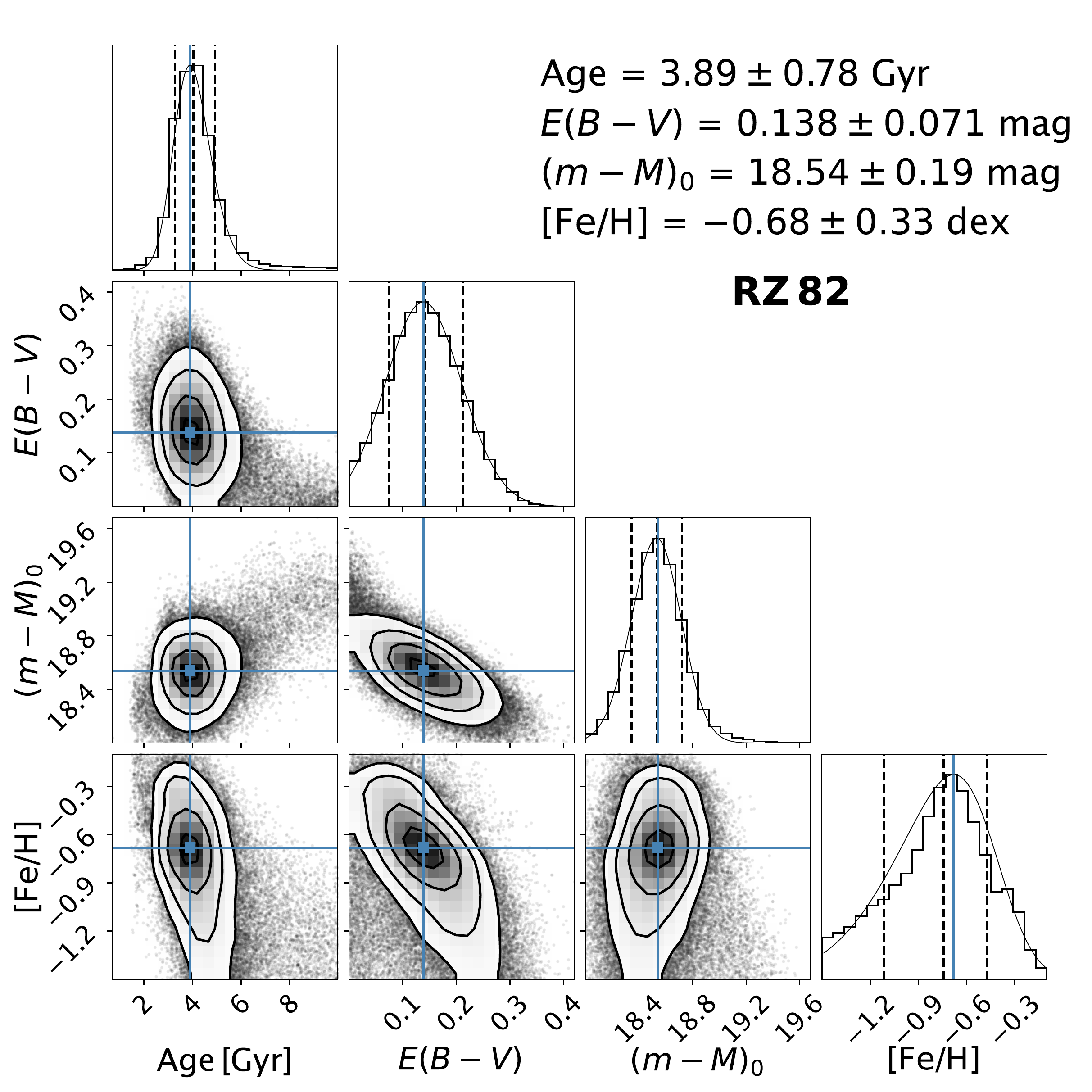}
    \includegraphics[width=0.33\textwidth]{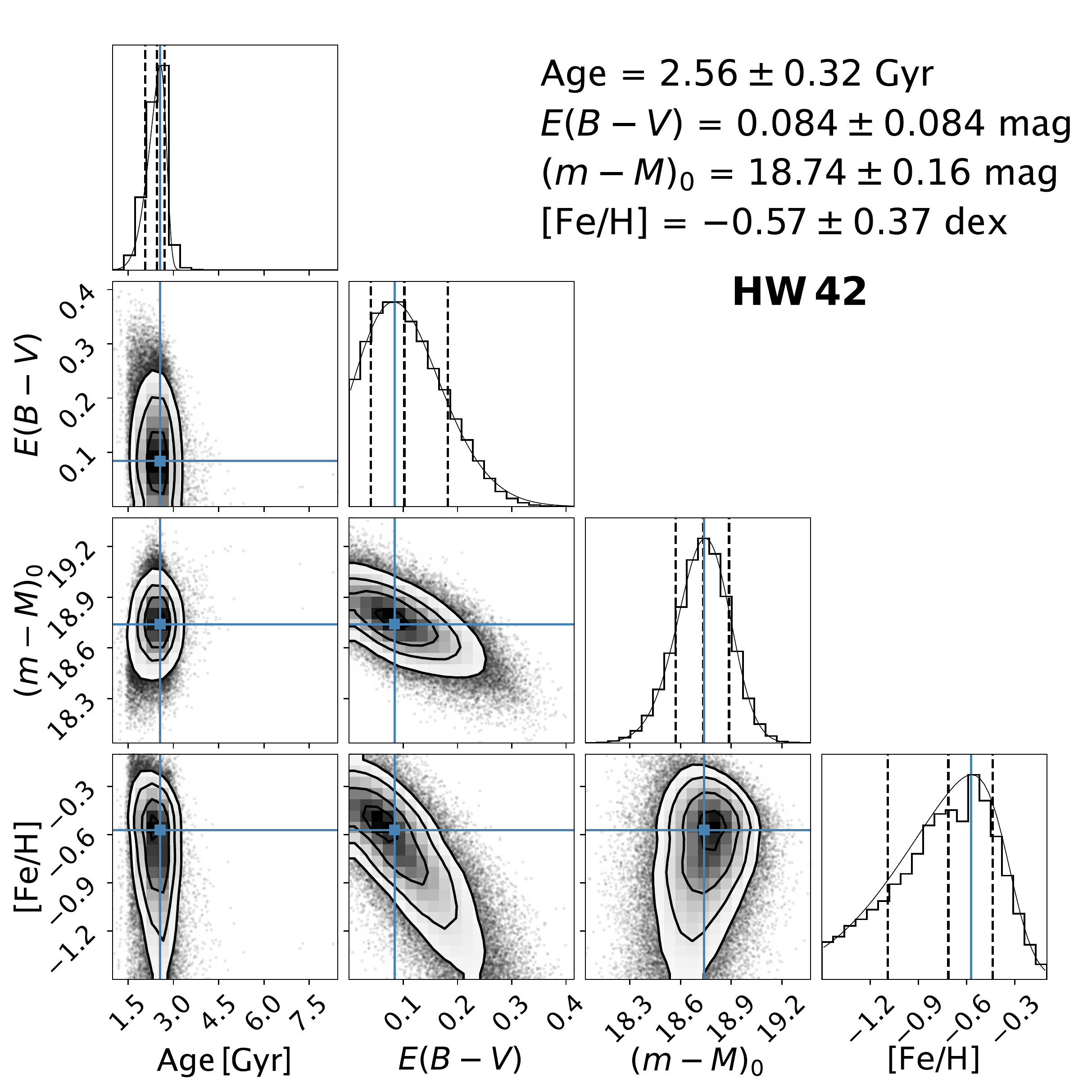}
    \includegraphics[width=0.33\textwidth]{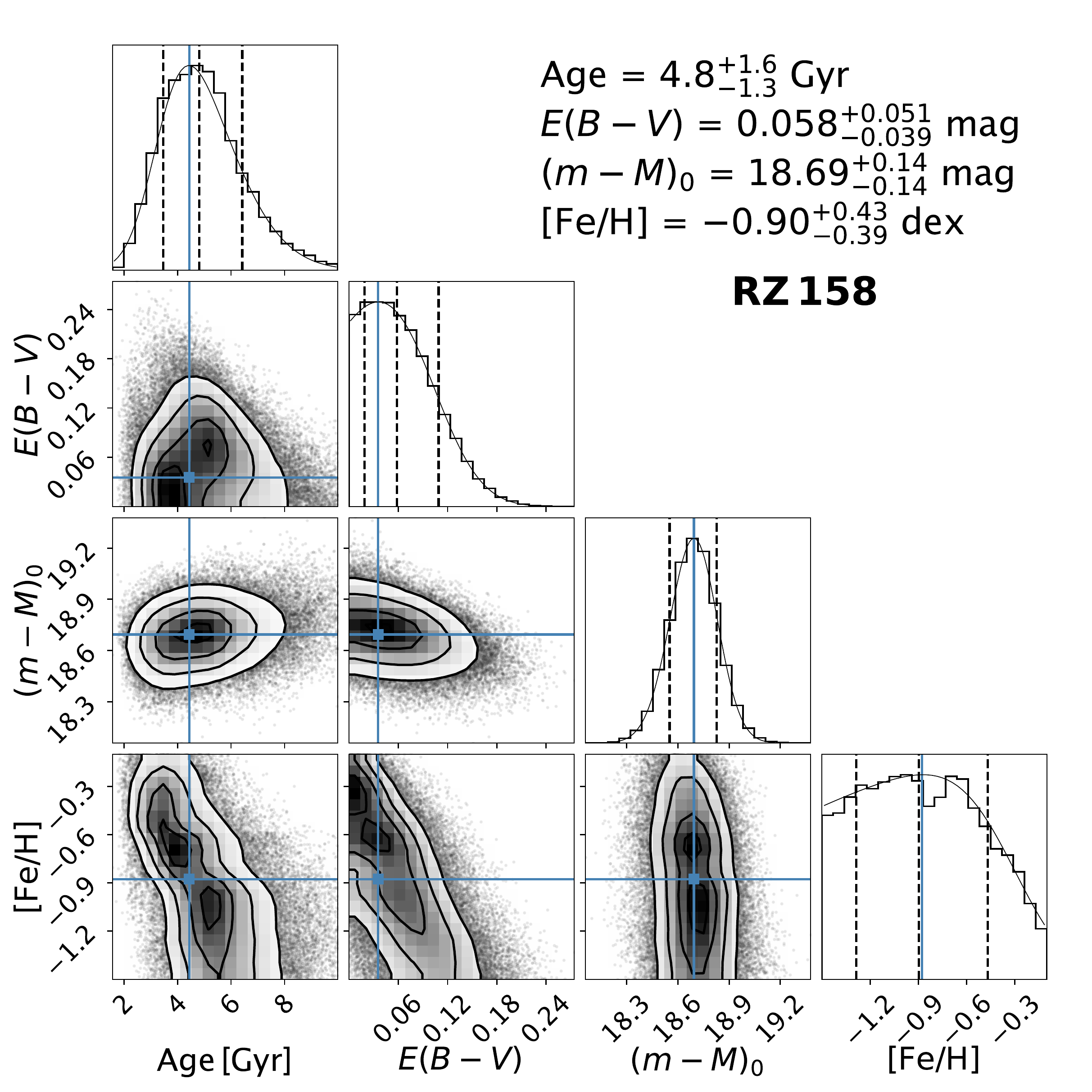}
    \caption{Corner plot of the resulting isochrone fits for the three sample clusters. The dashed lines in the histograms correspond to the median and the 16$^{\rm th}$ and 84$^{\rm th}$ percentiles as the $1\sigma$ levels. The contours in the 2D panels encompass the [$0.5\sigma$, $1.0\sigma$, $1.5\sigma$, $2.0\sigma$] levels. The posterior distributions in age shows that some walkers explore ages up to 10\,Gyr, but the convergence is obtained in lower ages.}
    \label{fig:cplot-all}
\end{figure*}

We employed the SIRIUS code \citep{skb20} to the decontaminated $V$ vs. $V-I$ CMDs, comparing the observed distribution of stars to theoretical PARSEC isochrones \citep{bmg12}. A geometrical likelihood was applied for each star in a Markov Chain Monte Carlo sampling, using the membership probability and the number of stars around it on the CMD as a uniform prior. The magnitude of the red clump (RC) stars was also adopted as a Gaussian prior in $(m-M)_0$, in order to match the RC of the isochrone according to the metallicity. Age, metallicity, distance, and reddening are free parameters during the fitting process. The SIRIUS code was previously employed to VISCACHA \citep[][hereafter \citetalias{dias+21}, \citetalias{dias+22}]{dias+21, dias+22} and \textit{HST} globular cluster data \citep[e.g.][]{kls19, osk20, skb20,souza+21}, as well as to e.g. VVV, \textit{Gaia} and 2MASS data \citep{trincado+21b,trincado+21a}. Table~\ref{tab:results} contains the derived parameters and $1\sigma$ uncertainties obtained from the posterior distributions.

Figure~\ref{fig:cmd-all} presents the decontaminated $V$ vs. $V-I$ CMDs of RZ\,82, HW\,42, and RZ\,158 with the best-fit isochrones and $1\sigma$ region. Member stars are colour-coded with the membership probability and field stars are shown in grey \citep{mcs10}. We also present an isochrone with the parameters from the literature \citep{rz05, p11c, ppv17}, showing that VISCACHA data rule out the possibility of an old age. Figure~\ref{fig:cplot-all} presents the respective corner plots, with the posterior distributions of the free parameters in the diagonal panels and the correlations between them in the other panels. 

The derived ages of $3.9$\,Gyr, $2.6$\,Gyr, and $4.8$\,Gyr for RZ\,82, HW\,42, and RZ\,158 are considerably younger than the previous literature values, thus redefining a brighter turnoff with the present deep, decontaminated VISCACHA photometry.
Nevertheless, the three clusters have a "Gyr morphology", with some blue stragglers and a red clump at $V\sim19.4$\,mag. The derived distances are suggestive of the location of the clusters relative to the SMC: $51.1$\,kpc for RZ\,82, which is projected in the foreground of the SMC bulk population \citep[$62$\,kpc;][]{gb15}, as illustrated by the crowded field in Fig. \ref{fig:ima}; $\sim 55$\,kpc for HW\,42 and RZ\,158, which are located in the Southern Bridge. Concerning the metallicity, the only spectroscopic value available to date is for HW\,42, for which \citet{debortoli+22} derived $\rm{[Fe/H]}=-0.58\pm0.03$ from CaT analysis, in excellent agreement with the photometric metallicity found in the present work. RZ\,158 shows a double peaked age-metallicity distribution in Fig. \ref{fig:cplot-all}. Preliminary spectroscopic metallicity of [Fe/H] = $-1.06\pm0.10$ dex (Dias et al. in prep.) supports the older peak around 5.5\,Gyr, and not the peak around 4.0 Gyr, which is still consistent with the conclusions of the present work.

%%%%%%%%%%%%%%%%%%%%%%%%%%%%%%%%%%%%%%%%%%%%%%%%%%%%%%
\section{Cluster masses}
\label{sect:mass}

We followed the procedures described in \cite[][hereafter \citetalias{santos+20}]{santos+20} to derive total mass for the clusters. In summary, we determined their integrated apparent $V$ magnitudes ($V_{int}$) by integrating the surface brightness profile from the centre out to the limiting radius, where the profile merges with the field stars surface brightness. We then converted $V_{int}$ to the absolute one ($M_V$) by using the clusters' individual distance and extinction from isochrone fitting (Table~\ref{tab:results}). Finally, the mass and its uncertainty was calculated following the calibration with age and metallicity (fixed at $Z=0.004$) of simple stellar population models given in \cite{mps14} and \citetalias{santos+20}. Mass uncertainty comes from propagation of errors in the measured surface brightness (propagated to $M_V$ error), age, extinction and distance. The integrated properties are shown in Table~\ref{tab:mass}.

\begin{table}
\caption{Integrated magnitudes and masses of investigated clusters.}
\label{tab:mass}
\centering
\begin{tabular}{lccc}
\hline
Cluster & $V_{int}$ & $M_V$ & $\log$(M/M$_\odot$) \\
\hline
RZ\,82 & $13.57\pm 0.17$ & $-5.39\pm 0.32$ & $4.05\pm 0.17$ \\
HW\,42 & $13.55\pm 0.09$ & $-5.45\pm 0.32$ & $3.97\pm 0.17$ \\
RZ\,158 & $14.21\pm 0.22$ & $-4.77\pm 0.30$ & $3.88\pm 0.18$ \\
\hline
\end{tabular}
\end{table}

%%%%%%%%%%%%%%%%%%%%%%%%%%%%%%%%%%%%%%%%%%%%%%%%%%
\section{Discussion}
\label{sec:dis}

The SMC clusters RZ\,82, HW\,42, and RZ\,158 analysed in this work are three of the oldest from the \citet{bica+20} catalogue, with ages of 7.1, 9.3, and 8.3\,Gyr old, respectively, based on integrated light or shallower photometry. We have shown that these clusters are actually 3.9, 2.6, and 4.8\,Gyr old, respectively, based on deeper and spatially resolved photometry. Some implications of the younger ages for these clusters are discussed below.

\begin{figure}
\centering
	\includegraphics[trim={0 0.6cm 0 1.9cm},clip,width=1.05\columnwidth]{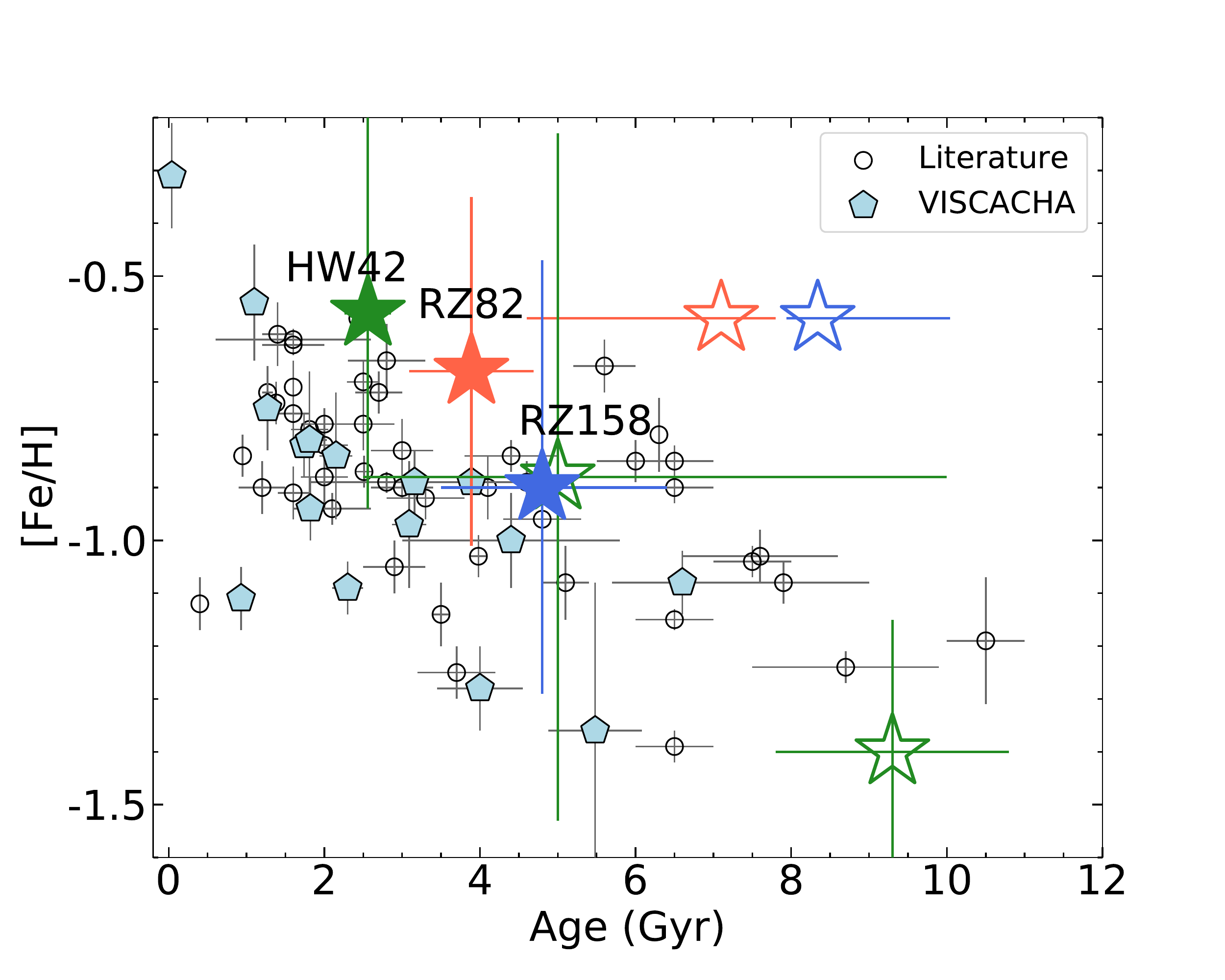}
    \caption{The age-metallicity relation of SMC star clusters. Black circles correspond to a literature compilation of homogeneous CaII triplet spectroscopic metallicities by \citet{dch98,parisi+09,parisi+15,p22,debortoli+22}, and ages from \citetalias{dias+21}; \citetalias{dias+22}; \citet{mighell+98,piatti+01,psc05,rz05,glatt+08,livanou+13,dkb14,parisi+14,li+16,nsc18,lagioia+19,narloch+21}. Previous VISCACHA results from \citetalias{mds19,dias+21,dias+22} are shown as blue pentagons. The filled stars are the clusters analysed in this work, whereas the empty stars are the same clusters but with the older ages and different metallicities from the literature (see text for details).}
    \label{fig:amr}
\end{figure}

The present clusters are consistent with the overall SMC enrichment history (e.g. \citetalias{dias+21}). Figure \ref{fig:amr} shows the age-metallicity relation of all SMC star clusters with available metallicities obtained with CaII triplet technique, all in the same scale, combined with ages from the best CMD with isochrone fitting available, some of them using \textit{HST} data (see compilation at \citealp{p22}). There is a large metallicity dispersion that is also seen in the multiple attempts of chemical evolution models to reproduce the SMC chemical evolution \citep[e.g.][]{debortoli+22}. The three clusters analysed in this work were supposedly among the oldest according to previous works (see compilation at \citealp{bica+20}). In particular, RZ\,82 and RZ\,158 had a combination of ages and metallicities that were off the bulk of SMC clusters and chemical enrichment models and the new ages and metallicities follow the trends now. HW\,42 had two previous determinations of age and metallicities, both in apparent agreement with the SMC evolution, which made it difficult for spotting any issue. Our new age and metallicity for this cluster says that it was formed more recently when the SMC was more metal-rich. In fact, our results are supported

by the spectroscopic metallicity. In summary, the age-metallicity relation of the SMC is sensitive to the diverse source of parameters, and it will be best traced by homogeneous and accurate parameters as provided by the VISCACHA survey. Moreover, all the SMC clusters older than $\sim 7$ Gyr are more metal-poor than [Fe/H]$\lesssim-1.0$.

The later evolution ($\tau \gtrsim 10^8$ yr) of star clusters involves the escape of stars by internal two body relaxation, commonly known as evaporation \citep[see e.g.][]{fall+09}. Low-mass stars are preferentially lost because the tendency that clusters have to reach energy equipartition combined with mass segregation, see e.g. \citet{bm03,kl08}. In this context, we have been searching for evolved star clusters that are missing low-mass stars, such as AM\,3 already identified in \citetalias{mds19}. 

\begin{figure} 
\centering
	\includegraphics[trim={0 0.5cm 0 2.2cm},clip,width=0.99\columnwidth]{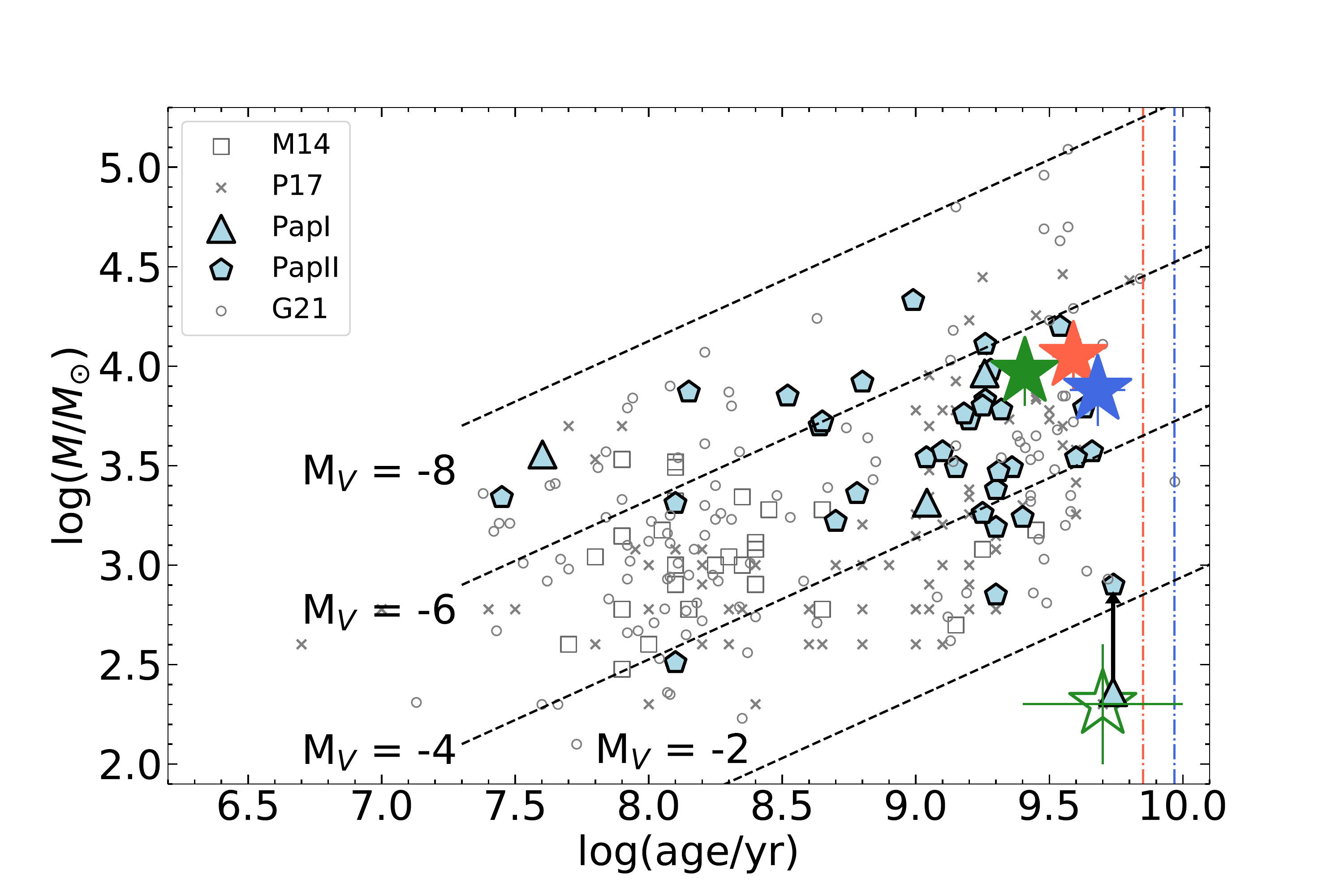}
    \caption{The age-mass relation of SMC star clusters. Literature data display the parameter space analysed by \citet[][M14]{mps14}, \citet[][P17]{ppv17}, \citetalias{mds19}, \citet[][PapII]{santos+20}, \citet[][G21]{gatto+21}. The iso-magnitude reference lines were adopted from M14 with metallicity [Fe/H]$\sim -0.7$. The blue triangle in the bottom right is AM\,3 with a lower limit mass from PapI and integrated mass from PapII indicated by the arrow. The three clusters analysed here are shown as filled stars as in Fig. \ref{fig:amr}. The empty star indicates the HW\,42 parameters by P17. The vertical blue and red lines show the ages for RZ\,82 and RZ\,158 that do not had published masses before.
    }
    \label{fig:dissolution}
\end{figure}

It has been argued that the evaporation of star clusters does not depend on their total mass \citep{chandar+10}. Nevertheless, the only cluster from the VISCACHA sample showing signs of dissolution so far is a relatively old and low-mass cluster, AM\,3 \citepalias{mds19}, i.e., an outlier in the age-mass parameter space (see Fig. \ref{fig:dissolution}. HW\,42 was previously analysed by \citet{ppv17}, with properties akin to AM\,3 in Fig.\ref{fig:dissolution}, which would be very interesting. However, our determination of a younger age by a factor of two and almost two orders of magnitude heavier mass places the cluster within the parameter space of the bulk of star clusters, not being an outlier anymore. RZ\,82 and RZ\,158 had age determinations placing them as outliers in Fig. \ref{fig:dissolution} regardless their mass, however, our new determination of ages and masses for these two clusters are also consistent with the bulk of intermediate-age SMC clusters.

%%%%%%%%%%%%%%%%%%%%%%%%%%%%%%%%%%%%%%%%%%%%%%%%%%%%%%%
\section{Conclusions}
\label{sec:conc}

We studied CMDs of the faint SMC clusters RZ\,82, HW\,42 and RZ\,158. The SOAR deep CMDs show well the turnoff regions, like in \citetalias{mds19}. The present ages of about $2-5$\,Gyr, are thus not as old as NGC\,121 which remains as the only SMC cluster with genuine CMD counterpart of Galactic globular clusters. The present clusters show metallicities of $\rm{[Fe/H]} \sim -0.60$ for RZ\,82 and HW\,42, and $\rm{[Fe/H]} =-0.90$ for RZ\,158. They fit well the cluster enrichment history of the SMC.

%%%%%%%%%%%%%%%%%%%%%%%%%%%%%%%%%%%%%%%%%%%%%%%%%%

\section*{Data availability}

The data underlying this article are available in the NOIRLab Astro Data Archive (https://astroarchive.noirlab.edu/) or upon reasonable request to the authors.

%%%%%%%%%%%%%%%%%%%%%%%%%%%%%%%%%%%%%%%%%%%%%%%%%%%%%%%
\section*{Acknowledgements}

The authors thank the anonymous referee for the helpful comments that improved the quality of the manuscript.
This study was financed in part by the Coordena\c c\~ao de Aperfei\c coamento de Pessoal de N\'ivel Superior - Brasil (CAPES) - Finance Code 001. F.Maia acknowledges financial support from Conselho Nacional de Desenvolvimento Científico e Tecnológico - CNPq (proc. 404482/2021-0) and from FAPERJ (proc. E-26/201.386/2022). R.A.P.O acknowledges FAPESP (proc. 2018/22181-0).  S.O.S. acknowledges FAPESP (proc. 2018/22044-3), the support of Deutsche Forschungsgemeinschaft (DFG, project number 428473034), and the DGAPA-PAPIIT grant IA103122. B.D. acknowledges support by ANID-FONDECYT iniciación grant No. 11221366.
Based on observations obtained at the Southern Astrophysical Research (SOAR) telescope, which is a joint project of the Minist\'erio da Ci\^encia, Tecnologia, e Inovação (MCTI) da Rep\'ublica Federativa do Brasil, the U.S. National Optical Astronomy Observatory (NOAO), the University of North Carolina at Chapel Hill (UNC), and Michigan State University (MSU).

%%%%%%%%%%%%%%%%%%%%%%%%%%%%%%%%%%%%%%%%%%%%%%%%%%%%%%%
\bibliographystyle{mnras}
\bibliography{bibliography}

\begin{thebibliography}{}
\makeatletter
\relax
\def\mn@urlcharsother{\let\do\@makeother \do\$\do\&\do\#\do\^\do\_\do\%\do\~}
\def\mn@doi{\begingroup\mn@urlcharsother \@ifnextchar [ {\mn@doi@}
  {\mn@doi@[]}}
\def\mn@doi@[#1]#2{\def\@tempa{#1}\ifx\@tempa\@empty \href
  {http://dx.doi.org/#2} {doi:#2}\else \href {http://dx.doi.org/#2} {#1}\fi
  \endgroup}
\def\mn@eprint#1#2{\mn@eprint@#1:#2::\@nil}
\def\mn@eprint@arXiv#1{\href {http://arxiv.org/abs/#1} {{\tt arXiv:#1}}}
\def\mn@eprint@dblp#1{\href {http://dblp.uni-trier.de/rec/bibtex/#1.xml}
  {dblp:#1}}
\def\mn@eprint@#1:#2:#3:#4\@nil{\def\@tempa {#1}\def\@tempb {#2}\def\@tempc
  {#3}\ifx \@tempc \@empty \let \@tempc \@tempb \let \@tempb \@tempa \fi \ifx
  \@tempb \@empty \def\@tempb {arXiv}\fi \@ifundefined
  {mn@eprint@\@tempb}{\@tempb:\@tempc}{\expandafter \expandafter \csname
  mn@eprint@\@tempb\endcsname \expandafter{\@tempc}}}

\bibitem[\protect\citeauthoryear{{Baumgardt} \& {Makino}}{{Baumgardt} \&
  {Makino}}{2003}]{bm03}
{Baumgardt} H.,  {Makino} J.,  2003, \mn@doi [\mnras]
  {10.1046/j.1365-8711.2003.06286.x}, \href
  {http://cdsads.u-strasbg.fr/abs/2003MNRAS.340..227B} {340, 227}

\bibitem[\protect\citeauthoryear{{Bica}, {Westera}, {Kerber}, {Dias}, {Maia},
  {Santos}, {Barbuy}  \& {Oliveira}}{{Bica} et~al.}{2020}]{bica+20}
{Bica} E.,  {Westera} P.,  {Kerber} L. d.~O.,  {Dias} B.,  {Maia} F.,  {Santos}
  Jr. J.~F.~C.,  {Barbuy} B.,   {Oliveira} R. A.~P.,  2020, \mn@doi [\aj]
  {10.3847/1538-3881/ab6595}, \href
  {https://ui.adsabs.harvard.edu/abs/2020AJ....159...82B} {159, 82}

\bibitem[\protect\citeauthoryear{{Bressan}, {Marigo}, {Girardi}, {Salasnich},
  {Dal Cero}, {Rubele}  \& {Nanni}}{{Bressan} et~al.}{2012}]{bmg12}
{Bressan} A.,  {Marigo} P.,  {Girardi} L.,  {Salasnich} B.,  {Dal Cero} C.,
  {Rubele} S.,   {Nanni} A.,  2012, \mn@doi [\mnras]
  {10.1111/j.1365-2966.2012.21948.x}, \href
  {http://adsabs.harvard.edu/abs/2012MNRAS.427..127B} {427, 127}

\bibitem[\protect\citeauthoryear{{Chandar}, {Fall}  \& {Whitmore}}{{Chandar}
  et~al.}{2010}]{chandar+10}
{Chandar} R.,  {Fall} S.~M.,   {Whitmore} B.~C.,  2010, \mn@doi [\apj]
  {10.1088/0004-637X/711/2/1263}, \href
  {https://ui.adsabs.harvard.edu/abs/2010ApJ...711.1263C} {711, 1263}

\bibitem[\protect\citeauthoryear{{Crowl}, {Sarajedini}, {Piatti}, {Geisler},
  {Bica}, {Clari{\'a}}  \& {Santos}}{{Crowl} et~al.}{2001}]{csp01}
{Crowl} H.~H.,  {Sarajedini} A.,  {Piatti} A.~E.,  {Geisler} D.,  {Bica} E.,
  {Clari{\'a}} J.~J.,   {Santos} Jr. J.~F.~C.,  2001, \mn@doi [\aj]
  {10.1086/321128}, \href {http://adsabs.harvard.edu/abs/2001AJ....122..220C}
  {122, 220}

\bibitem[\protect\citeauthoryear{{Da Costa} \& {Hatzidimitriou}}{{Da Costa} \&
  {Hatzidimitriou}}{1998}]{dch98}
{Da Costa} G.~S.,  {Hatzidimitriou} D.,  1998, \mn@doi [\aj] {10.1086/300340},
  \href {https://ui.adsabs.harvard.edu/abs/1998AJ....115.1934D} {115, 1934}

\bibitem[\protect\citeauthoryear{{De Bortoli}, {Parisi}, {Bassino}, {Geisler},
  {Dias}, {Gimeno}, {Angelo}  \& {Mauro}}{{De Bortoli}
  et~al.}{2022}]{debortoli+22}
{De Bortoli} B.~J.,  {Parisi} M.~C.,  {Bassino} L.~P.,  {Geisler} D.,  {Dias}
  B.,  {Gimeno} G.,  {Angelo} M.~S.,   {Mauro} F.,  2022, arXiv e-prints, \href
  {https://ui.adsabs.harvard.edu/abs/2022arXiv220515134D} {p. arXiv:2205.15134}

\bibitem[\protect\citeauthoryear{{De Grijs} \& {Bono}}{{De Grijs} \&
  {Bono}}{2015}]{gb15}
{De Grijs} R.,  {Bono} G.,  2015, \mn@doi [\aj] {10.1088/0004-6256/149/6/179},
  \href {https://ui.adsabs.harvard.edu/abs/2015AJ....149..179D} {149, 179}

\bibitem[\protect\citeauthoryear{{Dias}, {Kerber}, {Barbuy}, {Santiago},
  {Ortolani}  \& {Balbinot}}{{Dias} et~al.}{2014}]{dkb14}
{Dias} B.,  {Kerber} L.~O.,  {Barbuy} B.,  {Santiago} B.,  {Ortolani} S.,
  {Balbinot} E.,  2014, \mn@doi [\aap] {10.1051/0004-6361/201322092}, \href
  {https://ui.adsabs.harvard.edu/abs/2014A&A...561A.106D} {561, A106}

\bibitem[\protect\citeauthoryear{{Dias} et~al.,}{{Dias} et~al.}{2021}]{dias+21}
{Dias} B.,  et~al., 2021, \mn@doi [\aap] {10.1051/0004-6361/202040015}, \href
  {https://ui.adsabs.harvard.edu/abs/2021A&A...647L...9D} {647, L9}

\bibitem[\protect\citeauthoryear{{Dias} et~al.,}{{Dias} et~al.}{2022}]{dias+22}
{Dias} B.,  et~al., 2022, \mn@doi [\mnras] {10.1093/mnras/stac259}, \href
  {https://ui.adsabs.harvard.edu/abs/2022MNRAS.512.4334D} {512, 4334}

\bibitem[\protect\citeauthoryear{{Fall}, {Chandar}  \& {Whitmore}}{{Fall}
  et~al.}{2009}]{fall+09}
{Fall} S.~M.,  {Chandar} R.,   {Whitmore} B.~C.,  2009, \mn@doi [\apj]
  {10.1088/0004-637X/704/1/453}, \href
  {https://ui.adsabs.harvard.edu/abs/2009ApJ...704..453F} {704, 453}

\bibitem[\protect\citeauthoryear{{Fern{\'a}ndez-Trincado}
  et~al.,}{{Fern{\'a}ndez-Trincado} et~al.}{2021a}]{trincado+21b}
{Fern{\'a}ndez-Trincado} J.~G.,  et~al., 2021a, \mn@doi [\aap]
  {10.1051/0004-6361/202040255}, \href
  {https://ui.adsabs.harvard.edu/abs/2021A&A...647A..64F} {647, A64}

\bibitem[\protect\citeauthoryear{{Fern{\'a}ndez-Trincado}
  et~al.,}{{Fern{\'a}ndez-Trincado} et~al.}{2021b}]{trincado+21a}
{Fern{\'a}ndez-Trincado} J.~G.,  et~al., 2021b, \mn@doi [\apjl]
  {10.3847/2041-8213/abdf47}, \href
  {https://ui.adsabs.harvard.edu/abs/2021ApJ...908L..42F} {908, L42}

\bibitem[\protect\citeauthoryear{{Gatto} et~al.,}{{Gatto}
  et~al.}{2021}]{gatto+21}
{Gatto} M.,  et~al., 2021, \mn@doi [\mnras] {10.1093/mnras/stab2297}, \href
  {https://ui.adsabs.harvard.edu/abs/2021MNRAS.507.3312G} {507, 3312}

\bibitem[\protect\citeauthoryear{{Glatt} et~al.,}{{Glatt}
  et~al.}{2008}]{glatt+08}
{Glatt} K.,  et~al., 2008, \mn@doi [\aj] {10.1088/0004-6256/135/4/1106}, \href
  {https://ui.adsabs.harvard.edu/abs/2008AJ....135.1106G} {135, 1106}

\bibitem[\protect\citeauthoryear{{Glatt}, {Grebel}  \& {Koch}}{{Glatt}
  et~al.}{2010}]{ggk10}
{Glatt} K.,  {Grebel} E.~K.,   {Koch} A.,  2010, \mn@doi [\aap]
  {10.1051/0004-6361/201014187}, \href
  {https://ui.adsabs.harvard.edu/abs/2010A&A...517A..50G} {517, A50}

\bibitem[\protect\citeauthoryear{{Kerber} et~al.,}{{Kerber}
  et~al.}{2019}]{kls19}
{Kerber} L.~O.,  et~al., 2019, \mn@doi [\mnras] {10.1093/mnras/stz003}, \href
  {https://ui.adsabs.harvard.edu/abs/2019MNRAS.484.5530K} {484, 5530}

\bibitem[\protect\citeauthoryear{{Kruijssen} \& {Lamers}}{{Kruijssen} \&
  {Lamers}}{2008}]{kl08}
{Kruijssen} J.~M.~D.,  {Lamers} H.~J.~G.~L.~M.,  2008, \mn@doi [\aap]
  {10.1051/0004-6361:200810167}, \href
  {https://ui.adsabs.harvard.edu/abs/2008A&A...490..151K} {490, 151}

\bibitem[\protect\citeauthoryear{{Lagioia}, {Milone}, {Marino}  \&
  {Dotter}}{{Lagioia} et~al.}{2019}]{lagioia+19}
{Lagioia} E.~P.,  {Milone} A.~P.,  {Marino} A.~F.,   {Dotter} A.,  2019,
  \mn@doi [\apj] {10.3847/1538-4357/aaf729}, \href
  {https://ui.adsabs.harvard.edu/abs/2019ApJ...871..140L} {871, 140}

\bibitem[\protect\citeauthoryear{{Lasker} et~al.,}{{Lasker}
  et~al.}{2008}]{llm08}
{Lasker} B.~M.,  et~al., 2008, \mn@doi [\aj] {10.1088/0004-6256/136/2/735},
  \href {https://ui.adsabs.harvard.edu/abs/2008AJ....136..735L} {136, 735}

\bibitem[\protect\citeauthoryear{{Li}, {de Grijs}, {Deng}, {Geller}, {Xin},
  {Hu}  \& {Faucher-Gigu{\`e}re}}{{Li} et~al.}{2016}]{li+16}
{Li} C.,  {de Grijs} R.,  {Deng} L.,  {Geller} A.~M.,  {Xin} Y.,  {Hu} Y.,
  {Faucher-Gigu{\`e}re} C.-A.,  2016, \mn@doi [\nat] {10.1038/nature16493},
  \href {https://ui.adsabs.harvard.edu/abs/2016Natur.529..502L} {529, 502}

\bibitem[\protect\citeauthoryear{{Livanou}, {Dapergolas}, {Kontizas},
  {Nordstr{\"o}m}, {Kontizas}, {Andersen}, {Dirsch}  \& {Karampelas}}{{Livanou}
  et~al.}{2013}]{livanou+13}
{Livanou} E.,  {Dapergolas} A.,  {Kontizas} M.,  {Nordstr{\"o}m} B.,
  {Kontizas} E.,  {Andersen} J.,  {Dirsch} B.,   {Karampelas} A.,  2013,
  \mn@doi [\aap] {10.1051/0004-6361/201220926}, \href
  {https://ui.adsabs.harvard.edu/abs/2013A&A...554A..16L} {554, A16}

\bibitem[\protect\citeauthoryear{{Maia}, {Corradi}  \& {Santos}}{{Maia}
  et~al.}{2010}]{mcs10}
{Maia} F.~F.~S.,  {Corradi} W.~J.~B.,   {Santos} Jr. J.~F.~C.,  2010, \mn@doi
  [\mnras] {10.1111/j.1365-2966.2010.17034.x}, \href
  {http://adsabs.harvard.edu/abs/2010MNRAS.407.1875M} {407, 1875}

\bibitem[\protect\citeauthoryear{{Maia}, {Piatti}  \& {Santos}}{{Maia}
  et~al.}{2014}]{mps14}
{Maia} F.~F.~S.,  {Piatti} A.~E.,   {Santos} J.~F.~C.,  2014, \mn@doi [\mnras]
  {10.1093/mnras/stt2039}, \href
  {http://adsabs.harvard.edu/abs/2014MNRAS.437.2005M} {437, 2005}

\bibitem[\protect\citeauthoryear{{Maia} et~al.,}{{Maia} et~al.}{2019}]{mds19}
{Maia} F.~F.~S.,  et~al., 2019, \mn@doi [\mnras] {10.1093/mnras/stz369}, \href
  {http://adsabs.harvard.edu/abs/2019MNRAS.484.5702M} {484, 5702}

\bibitem[\protect\citeauthoryear{{Mighell}, {Sarajedini}  \&
  {French}}{{Mighell} et~al.}{1998}]{mighell+98}
{Mighell} K.~J.,  {Sarajedini} A.,   {French} R.~S.,  1998, \mn@doi [\aj]
  {10.1086/300591}, \href
  {https://ui.adsabs.harvard.edu/abs/1998AJ....116.2395M} {116, 2395}

\bibitem[\protect\citeauthoryear{{Narloch} et~al.,}{{Narloch}
  et~al.}{2021}]{narloch+21}
{Narloch} W.,  et~al., 2021, \mn@doi [\aap] {10.1051/0004-6361/202039623},
  \href {https://ui.adsabs.harvard.edu/abs/2021A&A...647A.135N} {647, A135}

\bibitem[\protect\citeauthoryear{{Nayak}, {Subramaniam}, {Choudhury}  \&
  {Sagar}}{{Nayak} et~al.}{2018}]{nsc18}
{Nayak} P.~K.,  {Subramaniam} A.,  {Choudhury} S.,   {Sagar} R.,  2018, \mn@doi
  [\aap] {10.1051/0004-6361/201732227}, \href
  {https://ui.adsabs.harvard.edu/abs/2018A&A...616A.187N} {616, A187}

\bibitem[\protect\citeauthoryear{{Oliveira} et~al.,}{{Oliveira}
  et~al.}{2020}]{osk20}
{Oliveira} R.~A.~P.,  et~al., 2020, \mn@doi [\apj] {10.3847/1538-4357/ab6f76},
  \href {https://ui.adsabs.harvard.edu/abs/2020ApJ...891...37O} {891, 37}

\bibitem[\protect\citeauthoryear{{Parisi}, {Grocholski}, {Geisler},
  {Sarajedini}  \& {Clari{\'a}}}{{Parisi} et~al.}{2009}]{parisi+09}
{Parisi} M.~C.,  {Grocholski} A.~J.,  {Geisler} D.,  {Sarajedini} A.,
  {Clari{\'a}} J.~J.,  2009, \mn@doi [\aj] {10.1088/0004-6256/138/2/517}, \href
  {https://ui.adsabs.harvard.edu/abs/2009AJ....138..517P} {138, 517}

\bibitem[\protect\citeauthoryear{{Parisi} et~al.,}{{Parisi}
  et~al.}{2014}]{parisi+14}
{Parisi} M.~C.,  et~al., 2014, \mn@doi [\aj] {10.1088/0004-6256/147/4/71},
  \href {https://ui.adsabs.harvard.edu/abs/2014AJ....147...71P} {147, 71}

\bibitem[\protect\citeauthoryear{{Parisi}, {Geisler}, {Clari{\'a}},
  {Villanova}, {Marcionni}, {Sarajedini}  \& {Grocholski}}{{Parisi}
  et~al.}{2015}]{parisi+15}
{Parisi} M.~C.,  {Geisler} D.,  {Clari{\'a}} J.~J.,  {Villanova} S.,
  {Marcionni} N.,  {Sarajedini} A.,   {Grocholski} A.~J.,  2015, \mn@doi [\aj]
  {10.1088/0004-6256/149/5/154}, \href
  {https://ui.adsabs.harvard.edu/abs/2015AJ....149..154P} {149, 154}

\bibitem[\protect\citeauthoryear{{Parisi}, {Gramajo}, {Geisler}, {Dias},
  {Clari{\'a}}, {Da Costa}  \& {Grebel}}{{Parisi} et~al.}{2022}]{p22}
{Parisi} M.~C.,  {Gramajo} L.~V.,  {Geisler} D.,  {Dias} B.,  {Clari{\'a}}
  J.~J.,  {Da Costa} G.,   {Grebel} E.~K.,  2022, arXiv e-prints, \href
  {https://ui.adsabs.harvard.edu/abs/2022arXiv220306542P} {p. arXiv:2203.06542}

\bibitem[\protect\citeauthoryear{{Perren}, {Piatti}  \& {V{\'a}zquez}}{{Perren}
  et~al.}{2017}]{ppv17}
{Perren} G.~I.,  {Piatti} A.~E.,   {V{\'a}zquez} R.~A.,  2017, \mn@doi [\aap]
  {10.1051/0004-6361/201629520}, \href
  {https://ui.adsabs.harvard.edu/abs/2017A&A...602A..89P} {602, A89}

\bibitem[\protect\citeauthoryear{{Piatti}}{{Piatti}}{2011}]{p11c}
{Piatti} A.~E.,  2011, \mn@doi [\mnras] {10.1111/j.1745-3933.2011.01105.x},
  \href {https://ui.adsabs.harvard.edu/abs/2011MNRAS.416L..89P} {416, L89}

\bibitem[\protect\citeauthoryear{{Piatti}, {Santos}, {Clari{\'a}}, {Bica},
  {Sarajedini}  \& {Geisler}}{{Piatti} et~al.}{2001}]{piatti+01}
{Piatti} A.~E.,  {Santos} J. F.~C.,  {Clari{\'a}} J.~J.,  {Bica} E.,
  {Sarajedini} A.,   {Geisler} D.,  2001, \mn@doi [\mnras]
  {10.1046/j.1365-8711.2001.04503.x}, \href
  {https://ui.adsabs.harvard.edu/abs/2001MNRAS.325..792P} {325, 792}

\bibitem[\protect\citeauthoryear{{Piatti}, {Santos}, {Clari{\'a}}, {Bica},
  {Ahumada}  \& {Parisi}}{{Piatti} et~al.}{2005}]{psc05}
{Piatti} A.~E.,  {Santos} Jr. J.~F.~C.,  {Clari{\'a}} J.~J.,  {Bica} E.,
  {Ahumada} A.~V.,   {Parisi} M.~C.,  2005, \mn@doi [\aap]
  {10.1051/0004-6361:20052982}, \href
  {http://adsabs.harvard.edu/abs/2005A%26A...440..111P} {440, 111}

\bibitem[\protect\citeauthoryear{{Piatti}, {Clari{\'a}}, {Bica}, {Geisler},
  {Ahumada}  \& {Girardi}}{{Piatti} et~al.}{2011}]{piatti+11}
{Piatti} A.~E.,  {Clari{\'a}} J.~J.,  {Bica} E.,  {Geisler} D.,  {Ahumada}
  A.~V.,   {Girardi} L.,  2011, \mn@doi [\mnras]
  {10.1111/j.1365-2966.2011.18627.x}, \href
  {https://ui.adsabs.harvard.edu/abs/2011MNRAS.417.1559P} {417, 1559}

\bibitem[\protect\citeauthoryear{{Rafelski} \& {Zaritsky}}{{Rafelski} \&
  {Zaritsky}}{2005}]{rz05}
{Rafelski} M.,  {Zaritsky} D.,  2005, \mn@doi [\aj] {10.1086/424938}, \href
  {https://ui.adsabs.harvard.edu/abs/2005AJ....129.2701R} {129, 2701}

\bibitem[\protect\citeauthoryear{{Santos} Jo{\~a}o F.~C. et~al.,}{{Santos}
  et~al.}{2020}]{santos+20}
{Santos} Jo{\~a}o F.~C. J.,  et~al., 2020, \mn@doi [\mnras]
  {10.1093/mnras/staa2425}, \href
  {https://ui.adsabs.harvard.edu/abs/2020MNRAS.498..205S} {498, 205}

\bibitem[\protect\citeauthoryear{{Souza}, {Kerber}, {Barbuy},
  {P{\'e}rez-Villegas}, {Oliveira}  \& {Nardiello}}{{Souza}
  et~al.}{2020}]{skb20}
{Souza} S.~O.,  {Kerber} L.~O.,  {Barbuy} B.,  {P{\'e}rez-Villegas} A.,
  {Oliveira} R.~A.~P.,   {Nardiello} D.,  2020, \mn@doi [\apj]
  {10.3847/1538-4357/ab6a0f}, \href
  {https://ui.adsabs.harvard.edu/abs/2020ApJ...890...38S} {890, 38}

\bibitem[\protect\citeauthoryear{{Souza} et~al.,}{{Souza}
  et~al.}{2021}]{souza+21}
{Souza} S.~O.,  et~al., 2021, \mn@doi [\aap] {10.1051/0004-6361/202141768},
  \href {https://ui.adsabs.harvard.edu/abs/2021A&A...656A..78S} {656, A78}

\bibitem[\protect\citeauthoryear{{Stetson}}{{Stetson}}{2000}]{s00}
{Stetson} P.~B.,  2000, \mn@doi [\pasp] {10.1086/316595}, \href
  {http://adsabs.harvard.edu/abs/2000PASP..112..925S} {112, 925}

\bibitem[\protect\citeauthoryear{{Tokovinin}, {Cantarutti}, {Tighe},
  {Schurter}, {Martinez}, {Thomas}  \& {van der Bliek}}{{Tokovinin}
  et~al.}{2016}]{tokovinin+16}
{Tokovinin} A.,  {Cantarutti} R.,  {Tighe} R.,  {Schurter} P.,  {Martinez} M.,
  {Thomas} S.,   {van der Bliek} N.,  2016, \mn@doi [\pasp]
  {10.1088/1538-3873/128/970/125003}, \href
  {https://ui.adsabs.harvard.edu/abs/2016PASP..128l5003T} {128, 125003}

\makeatother
\end{thebibliography}

\bsp	
\label{lastpage}

\end{document}